\documentclass[amsmath,amssymb,aps,pra,secnumarabic,superscriptaddress,showkeys,nobibnotes,twocolumn]{revtex4}

\usepackage{graphicx}
\usepackage{psfrag}

\begin{document}

\title{Top-Seeded Melt-Growth of YBa$_2$Cu$_3$O$_x$ Crystals for Neutron 
Diffraction Studies}

\author{D. C. Peets}
\email{dpeets@physics.ubc.ca}

\author{Ruixing Liang}
\altaffiliation{Canadian Institute for Advanced Research}
\affiliation{Department of Physics and Astronomy, University of British 
Columbia, Vancouver, BC, Canada}

\author{C. Stock}
\affiliation{Physics Department, University of Toronto, Toronto, ON, Canada}

\author{W. J. L. Buyers}
\altaffiliation{Canadian Institute for Advanced Research}

\author{Z. Tun}
\affiliation{National Research Council, Chalk River, ON, Canada}

\author{L. Taillefer}
\altaffiliation{Canadian Institute for Advanced Research}

\author{R. J. Birgeneau}
\altaffiliation{Canadian Institute for Advanced Research}
\affiliation{Physics Department, University of Toronto, Toronto, ON, Canada}

\author{D. A. Bonn}
\altaffiliation{Canadian Institute for Advanced Research}

\author{W. N. Hardy}
\altaffiliation{Canadian Institute for Advanced Research}
\affiliation{Department of Physics and Astronomy, University of British 
Columbia, Vancouver, BC, Canada}

\date{\today}

\begin{abstract}
We have grown cubic centimetre-size crystals of YBa$_2$Cu$_3$O$_x$ 
suitable for neutron studies, by a top-seeded melt-growth technique.  
Growth conditions were optimized with an eye toward maximizing phase 
purity.  It was found that the addition of 2\% Y$_2$BaCuO$_5$ and 
0.5\% Pt (by mass) were required to prevent melt loss and to obtain the 
highest crystallinity.  A neutron diffraction study on a mosaic of six 
such crystals found that the final Y$_2$BaCuO$_5$ concentration was 5\%, 
while other impurity phases comprised less than 1\% by volume.  The 
oxygen content was set to $x=6.5$, the crystals were detwinned and then 
carefully annealed to give the well-ordered ortho-II phase.  The neutron 
study determined that 70\% of the mosaic's volume was in the majority 
orthorhomic domain.  The neutron (0~0~6) and (1~1~0) rocking curve widths 
were $\sim$1$^\circ$ per crystal and $\sim$2.2$^\circ$ for the mosaic, and 
the oxygen chain correlation lengths were $>$100~\AA\ in the $a$- and 
$b$-directions and $\sim$50~\AA\ in the $c$-direction.  
\end{abstract}


\maketitle

\section{Motivation}

Neutron scattering is an extremely powerful technique for investigating 
spin fluctuation and magnetic ordering in superconducting samples.  
However, to obtain reasonable signal to noise one requires sample sizes on 
the order of several cubic centimetres.  The self-flux 
technique\cite{Erb,Ruixing}, which produces the highest-quality 
YBa$_2$Cu$_3$O$_x$ single crystals and is the primary method for growing 
crystals for fundamental research, has thus far only been able to yield 
crystals as large as 5~mm$\times$5~mm$\times\frac{1}{2}$~mm.  Additionally, 
YBa$_2$Cu$_3$O$_x$ is difficult to grow in an image furnace, due primarily 
to the low solubility of Y$_2$BaCuO$_5$ in the BaO-CuO melt\cite{diffuse}.  

The top-seeded melt-growth technique has been widely used to grow 
YBa$_2$Cu$_3$O$_x$ crystals as large as several 
inches\cite{old1,old2,old3,smg1,smg2,smg3,smg4,rod1}, but these crystals 
were intended for applications such as magnetic levitation.  The growth 
conditions were therefore optimized for maximal critical current 
densities, a criterion which requires defects for flux pinning.  As a 
result, the crystals typically contained 10-30\% Y$_2$BaCuO$_5$ and 
$\frac{1}{2}$-1\% Pt by mass, and were not well suited to neutron studies 
since the scattering from the inclusions can overlap the scattering from 
the YBa$_2$Cu$_3$O$_x$.  

This study aimed to optimize the top-seeded melt-growth technique for the 
lowest concentration of impurity phase inclusions and best crystallinity 
(minimum mosaic spread).  The resulting cubic centimetre-size 
YBa$_2$Cu$_3$O$_x$ crystals were then annealed to the oxygen-ordered 
ortho-II phase ($x=6.5$).  This allows a neutron scattering study on spin 
fluctuations and magnetic ordering to be carried out on higher-purity, 
well-ordered crystals.  

\section{Crystal Growth}
Precursor YBa$_2$Cu$_3$O$_x$ and Y$_2$BaCuO$_5$ powders were made by 
preparing stoichiometric mixtures of Y$_2$O$_3$ (99.999\% pure), BaCO$_3$ 
(99.999\%) and CuO (99.995\%), then calcining in BaZrO$_3$ crucibles at 
$\sim$950$^\circ$C until the reaction was complete as determined by X-ray 
diffraction.  The YBa$_2$Cu$_3$O$_x$, Y$_2$BaCuO$_5$ and Pt powders were 
mixed in an agate mortar under ethanol, then pressed in a cylindrical mold 
under 300~MPa of hydrostatic pressure.  The resulting pellet was 
approximately 13~mm high by 12~mm in diameter, and had a mass of 
$\sim$10~g.  

The addition of a small amount of Pt\cite{Pt} is required to prevent melt 
loss during growth by increasing the melt's viscosity.  We found that 
$\frac{1}{2}$\% Pt by mass was required for this purpose.  Without the 
addition of Y$_2$BaCuO$_5$, the crystals were porous, had poor mosaic 
spread, and contained BaO-CuO flux inclusions.  Best results were obtained 
when the pellet contained 2\% Y$_2$BaCuO$_5$ by mass.  

A higher concentration of Y$_2$BaCuO$_5$ was found to be necessary for 
proper seeding, so a small quantity of Y$_2$BaCuO$_5$-rich powder (10\% 
Y$_2$BaCuO$_5$ by mass) was added to the top of the pellet prior to
pressing.  This layer constituted $\sim$0.4\% of the pellet.  

\begin{figure}[htb]
\psfrag{t}{\small$t$}%
\psfrag{T}{\small\!\!$T_\circ$}%
\psfrag{slow}{\scriptsize$0.5\frac{^\circ\text{C}}{\text{h}}$}%
\psfrag{med}{\scriptsize$100\frac{^\circ\text{C}}{\text{h}}$}%
\psfrag{fast}{\scriptsize$200\frac{^\circ\text{C}}{\text{h}}$}%
\psfrag{990}{\scriptsize$990^\circ$C, 16h}%
\psfrag{1030}{\scriptsize$1030^\circ$C, 4h}%
\psfrag{975}{\scriptsize$975^\circ$C}%
\psfrag{grad}{}%
\psfrag{nograd}{\scriptsize Gradient}%
\psfrag{stillnograd}{\scriptsize applied}%
\psfrag{BaZrO3}{\scriptsize BaZrO$_3$}%
\psfrag{Seed}{\scriptsize NdBa$_2$Cu$_3$O$_x$ seed}%
\psfrag{2
\psfrag{Y211-rich}{\scriptsize Y$_2$BaCuO$_5$--}%
\psfrag{layer}{\scriptsize rich layer}%
\includegraphics[width=0.95\columnwidth]{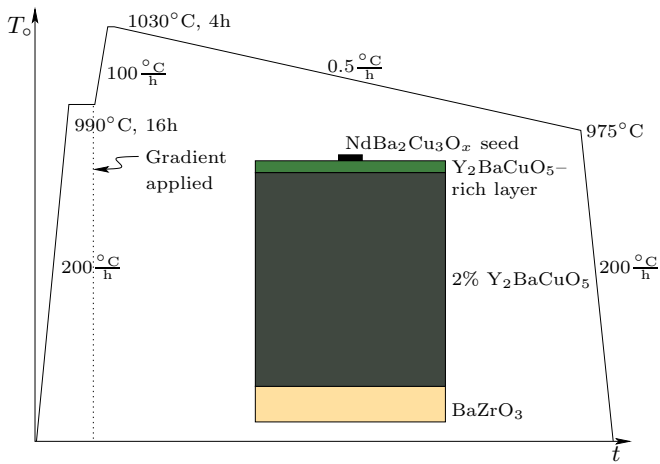}\hspace{0.025\columnwidth}
\caption{\label{fig:pellet}The crystallization furnace program, where 
$T_\circ$ is the average temperature of the pellet and $t$ is the time.  
The inset depicts the growth setup, showing the geometry of the pellet, 
seed and substrate.}
\end{figure}

NdBa$_2$Cu$_3$O$_x$ was chosen as a seed crystal for its near-perfect 
lattice match and for having a melting temperature $\sim$80$^\circ$C 
higher than that of YBa$_2$Cu$_3$O$_x$.  The seed and pellet were placed 
on a disc-shaped substrate, loaded into a three-zone vertical tube 
furnace, and subjected to the temperature program shown in figure 
\ref{fig:pellet}.  After a sintering step at 990$^\circ$C, a temperature 
gradient of $\sim5\frac{^\circ\text{C}}{\text{cm}}$ was applied and the 
pellet was melted, then slowly cooled.  This effectively moved the 
peritectic temperature through the pellet, from seed to substrate, over 
the course of several days.  

While several substrates were tested, including alumina (Al$_2$O$_3$) and 
single-crystalline MgO and SrTiO$_3$, satisfactory results were only 
obtained using BaZrO$_3$ discs.  

The top of each crystal had an elevated Y$_2$BaCuO$_5$ concentration and 
often extraneous domains near the edge, while the bottom had a layer with 
a high concentration of impurity phases which had been pushed there by the 
growth front (including much of the platinum).  The top and bottom 
millimetres of the crystal were accordingly cut off using a diamond saw.  
Additionally, a millimetre was removed from two opposite sides, to create 
flat (1~0~0)/(0~1~0) faces for detwinning.  We were able to align the 
crystals visually to within a degree.  

To form the oxygen-ordered ortho-II phase of YBa$_2$Cu$_3$O$_{6.5}$, 
which has alternating full and empty chains, the oxygen content was set to 
$x=6.53$ by annealing for one week in 1~atm oxygen gas, at 760.0$^\circ$C, 
in the same conditions as for single crystals\cite{OII}.  It has been 
reported that the ortho-II phase's longest oxygen correlation lengths may 
be realized at oxygen contents slightly above 6.50\cite{OII2}.  

\begin{figure}[htb]
\psfrag{Element}{\scriptsize Furnace element}%
\psfrag{Insulation}{\scriptsize Insulation}%
\psfrag{Nitrogen}{\scriptsize N$_2$ flow}%
\includegraphics*[width=\columnwidth]{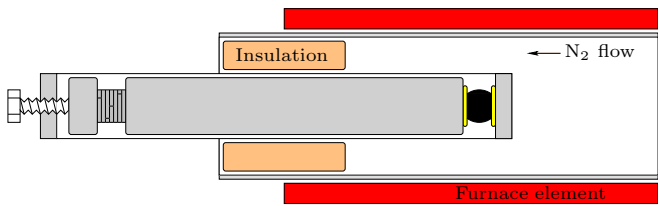}
\caption{\label{fig:detwinner}Schematic diagram of the detwinner.}
\end{figure}

A specialized detwinner, depicted schematically in figure 
\ref{fig:detwinner}, was built for these crystals.  The sample was loaded 
into it, sandwiched between two gold pads to ensure that the pressure was 
uniformly distributed on the two $a$-axis faces.  A uniaxial pressure of 
$\sim$150~MPa (1500~atm) was applied, as gauged by monitoring the 
compression of several lock washers with known spring constants.  The 
detwinner was loaded into a horizontal tube furnace and heated to 
350$^\circ$C in flowing nitrogen for 30 minutes, then cooled to room 
temperature at 60$^\circ$C per hour.  The crystal's mass was not observed 
to be changed by this process, indicating no measurable change in the 
crystal's oxygen content during detwinning.  

Once detwinned, the crystals were annealed at 60$^\circ$C in sealed 
containers for two weeks to establish ortho-II oxygen ordering.  

\section{Analysis}
Magnetization measurements in a Quantum Design SQUID magnetometer found 
the crystals' $T_c$ to be 59~K, and 2.5~K wide (field-cooled, $H=1.5$~Oe, 
$\vec H\parallel\vec c$).  

A sample was subjected to EDX compositional analysis to determine the 
shape and distribution of impurity phases.  Y$_2$BaCuO$_5$ was only 
observed at the base of the pellet, near the growth front, but the 
equipment's $\sim1\mu$m resolution makes it insensitive to small 
inclusions in the bulk.  Two Pt-containing phases were found, one of which 
was consistent with the formula Y$_2$PtBa$_3$Cu$_2$O$_{10}$.  Neutron 
scans found weak impurity peaks consistent with this compound as well.  

\begin{figure}[htb]
\includegraphics[width=\columnwidth]{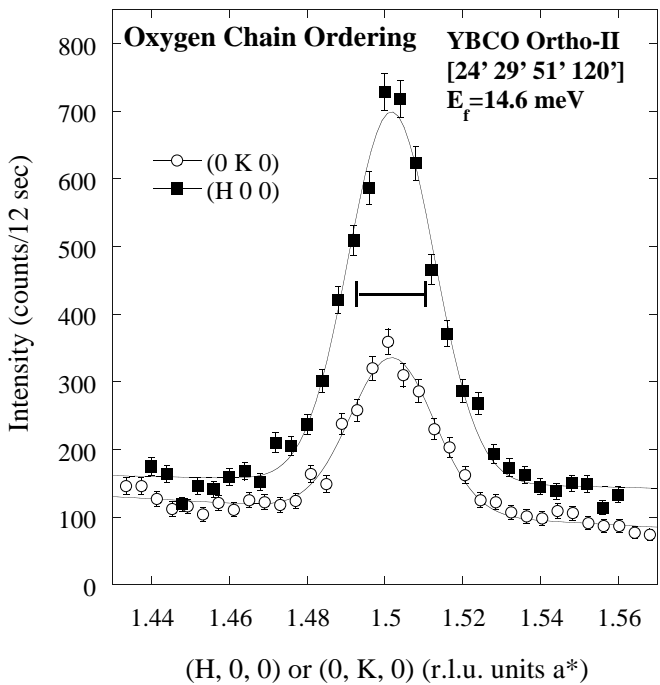}
\caption{\label{fig:neutron}Radial elastic neutron scattering scans 
through ($\frac{3}{2}a^*$~0~0) and (0~$\frac{3}{2}a^*$~0), with the 
experimental resolution indicated.  The larger and smaller peaks are from 
the majority and minority domains, the larger domain accounting for 70\% 
of the sample volume.}
\end{figure}

Neutron diffraction experiments were carried out at the NRC Chalk River 
laboratory (NRU reactor).  There, six crystals were sealed in airtight 
cans under dry helium, aligned, and inserted into the E3 spectrometer.  
The (0~0~6) and (1~1~0) rocking curve widths were $\sim$1$^\circ$ for each 
crystal, and $\sim$2.2$^\circ$ for the mosaic of six.  Figure \ref{fig:neutron} 
shows ortho-II ordering superlattice diffractions via radial scans through 
($\frac{3}{2}a^*$~0~0) and (0~$\frac{3}{2}a^*$~0).  The detwinning was 
incomplete -- the majority domain occupied only 70\% of the sample volume.  
The neutron study also found that the concentration of Y$_2$BaCuO$_5$ in 
the mosaic was $\sim$5\% by volume, while all other impurity phases 
constituted less than 1\%.  

A determination of the oxygen ordering correlation length was resolution 
limited (horizontal bar in figure \ref{fig:neutron}), but a fit to a 
resolution-convolved Lorentzian indicated a length $>$100~\AA\ in the 
$a$- and $b$-directions and $\sim$50~\AA\ in the $c$-direction.  Given the 
growth method used, these compare remarkably well with the values of 
$\xi_a=148$~\AA, $\xi_b=430$~\AA\ and $\xi_c=58$~\AA\ obtained in 
high-purity single crystals\cite{OII}.  Indeed, they even exceed the 
correlation lengths found in many flux-grown single crystals.  

Further neutron scattering results will be published 
elsewhere\cite{stock2002}.  

\section{conclusion}
We have grown and detwinned cubic centimetre-size crystals of ortho-II 
YBa$_2$Cu$_3$O$_{6.5}$, with oxygen ordering correlation lengths $>$100~\AA\ 
in the plane and $\sim$50~\AA\ in the $c$-direction.  They contain 5\% 
Y$_2$BaCuO$_5$ by volume and are partially detwinned, with 70\% of their 
volume in the major orientation.  Each crystal has a rocking curve width 
of $\sim$1$^\circ$.  These crystals are currently being studied in detail 
by neutron scattering, and have already shown that their symmetry is not 
broken by satellite Bragg peaks associated with static long-range 
d-density wave order, as is seen in less-ordered crystals\cite{stock2002}.  

\begin{acknowledgments}
This research was supported by the Canadian Institute for Advanced 
Research.  The work at the University of British Columbia and at the 
University of Toronto was supported by NSERC.  
\end{acknowledgments}

\bibliography{smg-ybco}

\end{document}